# The role of water in the preparation and stabilization of high quality phosphorene flakes

Manuel Serrano-Ruiz,[a] Maria Caporali,[a,]* Andrea Ienco,[a] Vincenzo Piazza,[b] Stefan Heun,[c] Maurizio Peruzzini[a,]*

[a] Istituto di Chimica del Composti Organometallici, Consiglio Nazionale delle Ricerche (CNR-ICCOM), Via Madonna del Piano 10, 50019 Sesto Fiorentino, Italy

[b] Center for Nanotechnology Innovation @NEST, Istituto Italiano di Tecnologia, Piazza San Silvestro 12, I-56127 Pisa, Italy

[c] NEST, Istituto Nanoscienze-CNR and Scuola Normale Superiore, Piazza S. Silvestro 12, Pisa, Italy.

*This paper is dedicated to Prof. Dr. Manfred Scheer on the occasion of his 60$^{th}$ birthday*

E-mail: maurizio.peruzzini@iccom.cnr.it

Since the first reports by P.D. Ye[1] and Y. Zhang[2] on the micromechanical exfoliation of bulk layered black phosphorus (BP) were published in January 2014,[3] more than two hundred papers have been published up to today, dealing with experimental and mostly theoretical studies on phosphorene, the two-dimensional BP monolayer.[4] The so called "renaissance" of black phosphorus arises from the unique properties of this material, endowed with a direct band gap of 0.3 eV in the bulk which increases to 2.0 eV going down to the monolayer. In addition to its intrinsic semiconducting properties, its puckered structure[5] gives anisotropic in-plane properties that make BP and its 2D derivative, phosphorene, very promising candidates for nanoelectronics and nanophotonics applications.[6] Therefore, the preparation of high quality single and few-layer nanosheets of BP has attracted enormous interest in the scientific community. First the micromechanical cleavage (Scotch tape method) was applied successfully as mentioned above, then the liquid exfoliation[7] and the electrochemical exfoliation[8] of black phosphorus (BP) were explored as well. Several groups were working simultaneously and independently on the liquid exfoliation of BP under sonication. Therefore, distinct contributions came out in a relatively short time.[7] In particular, Hersam[7c] and Salehi-Khojin[7d] performed successfully the liquid exfoliation of BP in dimethylsulfoxide (DMSO) and dimethylformamide, both having a high dielectric constant and a high surface tension which are prerequisite. Molecular dynamics simulations showed that the cohesive energy between the solvent molecules and BP layer is very important and explains why DMSO, having a strong adhesion with phosphorene, assures a great stability of the dispersion in the time.[9] A recent paper by S.C. Warren *et al.*[7e] surveyed 18 organic solvents for their ability to exfoliate black phosphorus. A successful scale-up of the exfoliation process of BP at the 10 g scale was achieved using N-methyl-2-pyrrolidone (NMP). Ultrathin BP nanosheets were prepared by Xie[7f] sonicating grinded BP in water, the substitution of an organic solvent with water was highly relevant and allowed the authors to test the resultant BP suspension in biological applications. In our laboratories, the liquid exfoliation of BP in dimethylsulfoxide (DMSO) was repeated analyzing carefully the amount of water present in DMSO, being this a highly hygroscopic solvent which always contains traces of water, even after treatment with drying agents. Surprisingly, water resulted to be not an innocent player in the process under study. We therefore performed a systematic study evaluating the influence of the amount of water on the exfoliation, stability of the resulting suspension, and quality of phosphorene flakes.

Highly pure and crystalline BP was prepared following a published procedure.[10] The liquid exfoliation was carried in DMSO containing $H_2O$ using an ultrasonication bath working at 37 KHz frequency, while keeping the suspension under nitrogen in a sealed glass vial and in the dark. Recently some authors have outlined the importance of anaerobic and dark environment for both the mechanical and liquid exfoliation of BP nanosheets, that can irreversibly degrade to oxidized compounds.[11] In particular, the environmental instability of mechanically exfoliated few-layer BP was investigated,[10a] but the influence of the amount of water in the liquid exfoliation process has not yet been analyzed in detail.

As shown in Scheme 1, three different ranges of molar ratio between black phosphorus and water were studied: a) molar ratio (P/$H_2O$) ≥ 15, b) molar ratio (P/$H_2O$) between 14 and 1.5; c) molar ratio (P/$H_2O$) between 1.4 and 0.3. The limits chosen are a compromise mainly between the stability of exfoliated BP, the stability of the suspensions and the quality of flakes. For instance, the product of the exfoliation of BP with a molar ratio of 15 already decomposes during the sonication experiment. On the opposite side, when a molar ratio P/$H_2O$ less of 1.4 was used, a bad quality material was obtained with most flakes' thickness greater than 100 nm. Finally, in the intermediate P/$H_2O$ range (between 14 and 1.5) "good" quality phosphorene flakes, in terms of stability of the suspension and thickness of the exfoliated flakes, may always be obtained.

In case a, we dealt with adventitious water, usually present in commercially available dry DMSO and the relative content was properly measured by $^1$H NMR, while in cases b) and c) a specific amount of water was added. After sonication for a proper period of time, three different suspensions were obtained, which looked quite different as shown in Scheme 1. They were analysed by $^1$H, $^{31}$P{$^1$H} and $^{31}$P NMR at regular intervals of time, each time allowing the suspension to stand by for three hours, performing the measurements, and then proceeding with the ultrasonication.

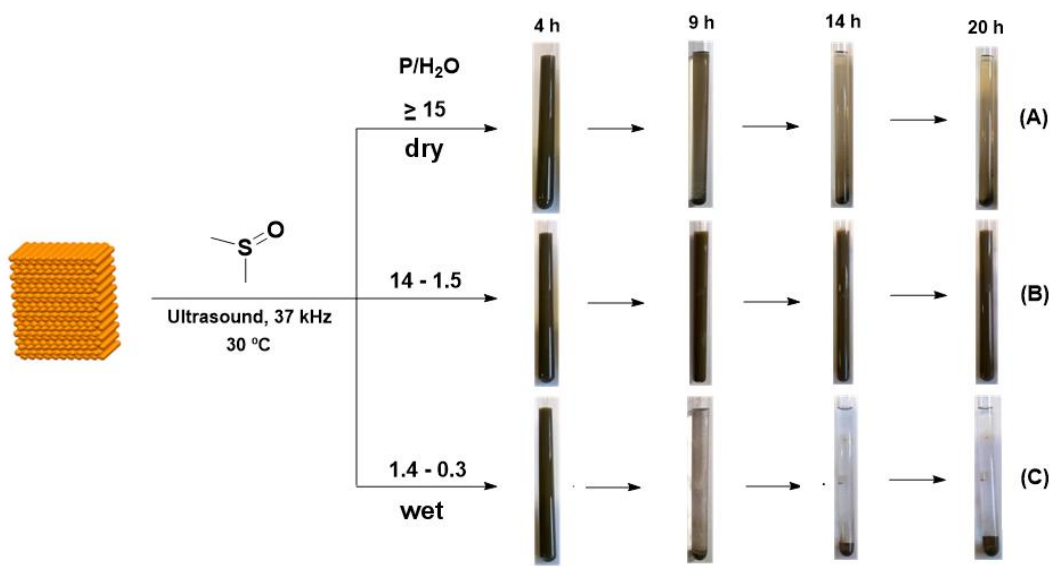

Scheme 1. Influence of water amount in the liquid exfoliation process of BP in DMSO at different time (4, 9, 14, 20h).

After 20 hours of sonication, the characterization of the suspensions of BP nanoflakes was carried out by drop-casting the sol on a suitable support, in the case of AFM, SEM and Raman onto a 300 nm silicon dioxide on silicon wafers, or onto a carbon copper grid for TEM, leaving the suspension in contact with the support for two min to allow for the formation of a film, before washing with acetone, and drying with a dry nitrogen flow.

Test A) P/H$_2$O ≥ 15

$^{31}$P{$^1$H} NMR was measured in the course of exfoliation and showed in the beginning the formation of species with resonances between +6 ppm and -13 ppm, which

gradually disappeared, see Figure 1. After 14 hours of sonication, the solution looked clear and yellow, while the initial amount of BP was completely converted as evidenced by the disappearance of the dark grey solid while $^{31}P\{^1H\}$ NMR showed the presence of a soluble species having a featureless singlet at -24.9 ppm, see top spectrum in Figure 1.

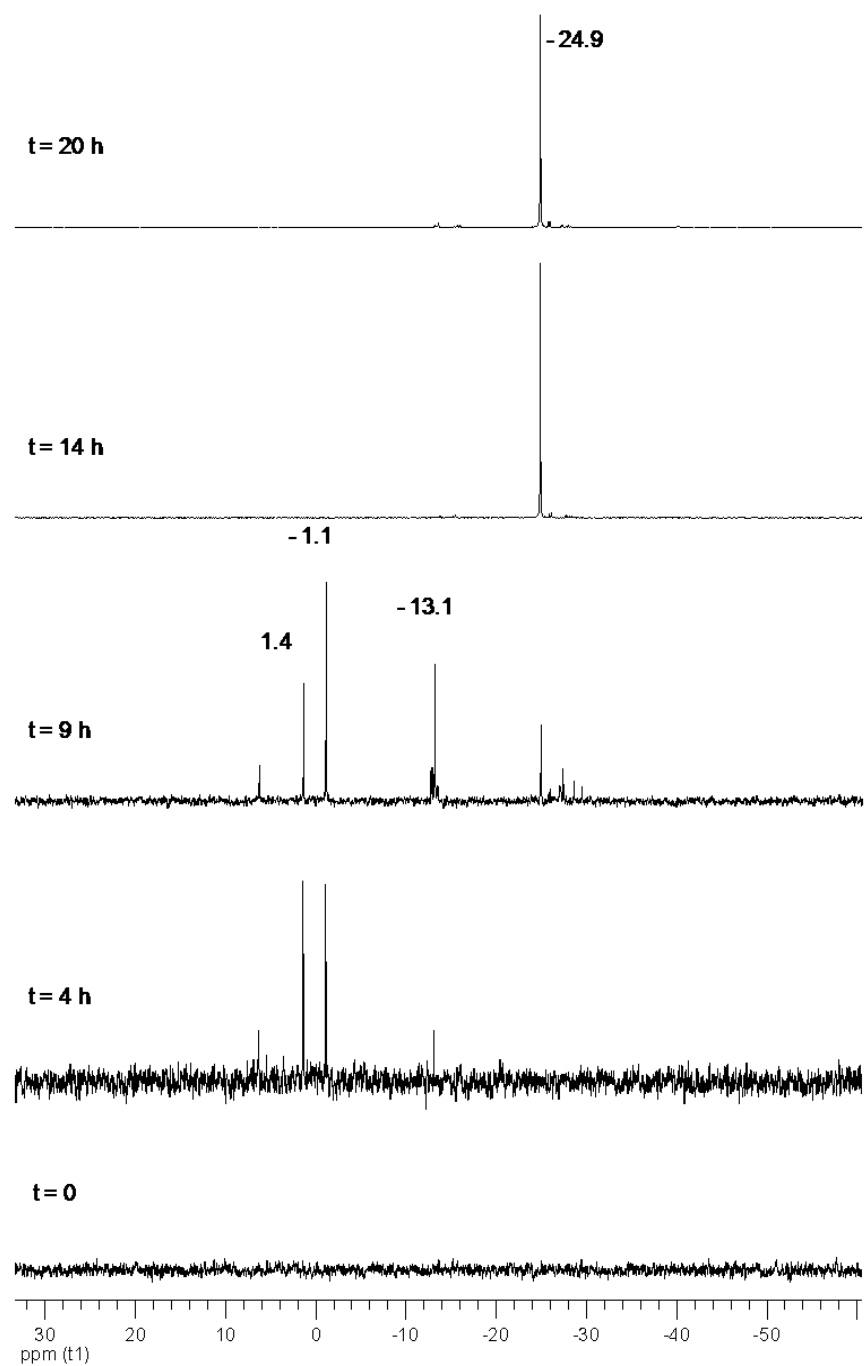

Figure 1. $^{31}P\{^1H\}$ NMR (DMSO-$d_6$, T = 25 °C, 121.49 MHz) of Test A. The vertical scale of the spectra taken at t = 4h and t = 9h is emphasized in order to show the signals due to first degradation products.

The resonances at 1.4 ppm and -1.1 ppm, which firstly arose in the $^{31}$P spectrum, were unequivocally assigned to phosphorous acid, $H_3PO_3$, and phosphoric acid, $H_3PO_4$, respectively, by adding to the sonicated sample (4 h), a small amount of these oxyacids (see also figures S1 and S2).[12]

Noticeably, the formation of $H_3PO_3$ as degradation product in the liquid exfoliation of BP has been recently predicted by Coleman and co-workers[13] who theoretically investigated the water degradation of few-layer BP. They anticipated that phosphorene disproportionation should generate also $PH_3$, which however was never detected in our experiments

The other degradation products resonating at - 13.1 and -24.9 ppm were assigned to pyrophosphate, $[HP_2O_6]^-$, and to trimetaphosphate $[H_2P_3O_9]^-$ respectively, on the basis of high resolution ESI MS (Figure 2) and by comparison with mass spectral data of these compounds.[14] The peaks at m/z = 256.90217, m/z = 176.93592, and m/z = 158.92544 can be attributed to derivatives coming from the hydrolysis of trimetaphosphate.[15]

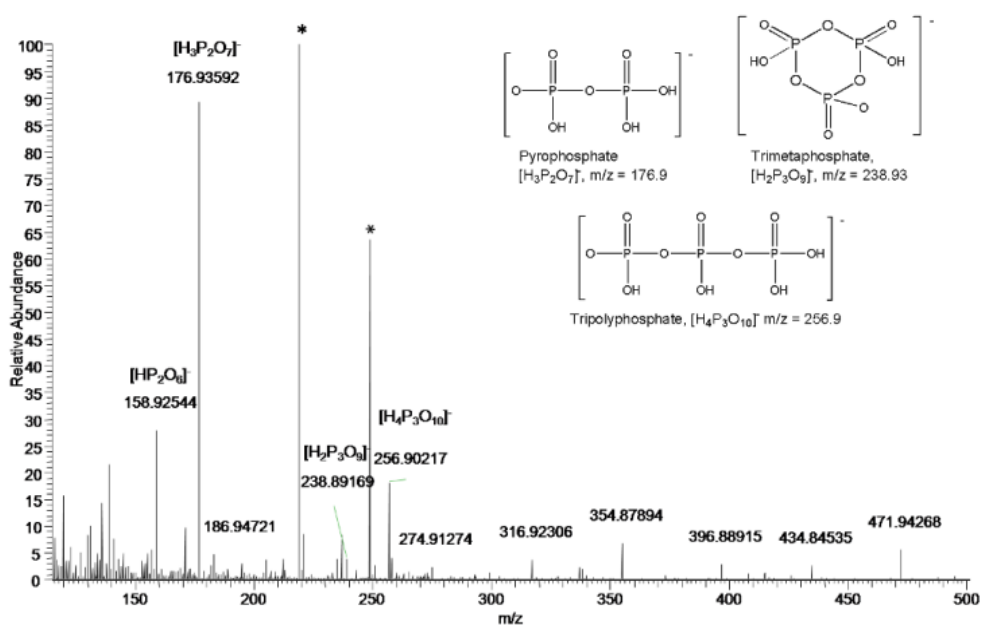

Figure 2. High resolution ESI MS on the reaction mixture acquired in negative ion mode. The peaks labeled with a star are impurities due to the background. The sketches of the polyphosphate anions are depicted in the inset.

Although further mechanistic studies are needed and further experiments are in progress, we can hypothesize that the formation of polyphosphate anions during the BP exfoliation in anhydrous DMSO originates from DMSO itself. In agreement with this hypothesis, the $^1$H NMR spectrum carried out after prolonged sonication (20 h), showed, apart from a singlet at δ = 14.1, (Figure S3), which is typical for the acidic proton of polyphosphates, a singlet at 2.6 ppm, which was attributed to dimethylsulfide.[16] The presence of $SMe_2$ implicates that DMSO may effectively transfer oxygen atoms to phosphorus during the degradation of exfoliated phosphorene sheets in anhydrous DMSO. This intriguing and unexpected reaction mechanism is currently under study in our group by means of theoretical calculations and selected NMR tests on specific samples (including different allotropes of elemental phosphorus).

In practice, when DMSO is "really" anhydrous, the exfoliation of BP is likely complete but as soon as BP is exfoliated, most of the nanosheets react with the solvent under the current ultrasound conditions, causing an extended cleavage of the P-P bond network, which eventually results in the formation of degradation products as discussed above. In line with this hypothesis, no BP atomic layers were observed by optical microscopy, AFM and TEM, but only drops of the solvent, see Figure S6 and S7, respectively.

Test B) 1.5 < $P/H_2O$ <14

Working with a $P/H_2O$ molar ratio higher than 1.5 yielded a brown suspension. After 20 h sonication, centrifugation at 6000 rpm for one hour left a grey solid residue and a light yellow supernatant (see Figure 3).

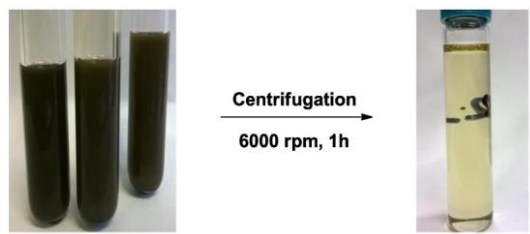

Figure 3. Phosphorene dispersion in DMSO before and after centrifugation ($P/H_2O$ = 2).

After sedimentation of the heavier flakes, inspection of the yellow supernatant with optical microscope, AFM, micro-Raman, and TEM revealed that the dark residue is constituted by BP flakes having a high number of atomic layers (*i.e.* height profile from 30 nm to 100 nm), while the yellow supernatant contains more homogeneous and "lighter" phosphorene flakes having few atomic layers, with an average size 500 nm x 800 nm, see TEM image in Figure 4, left.

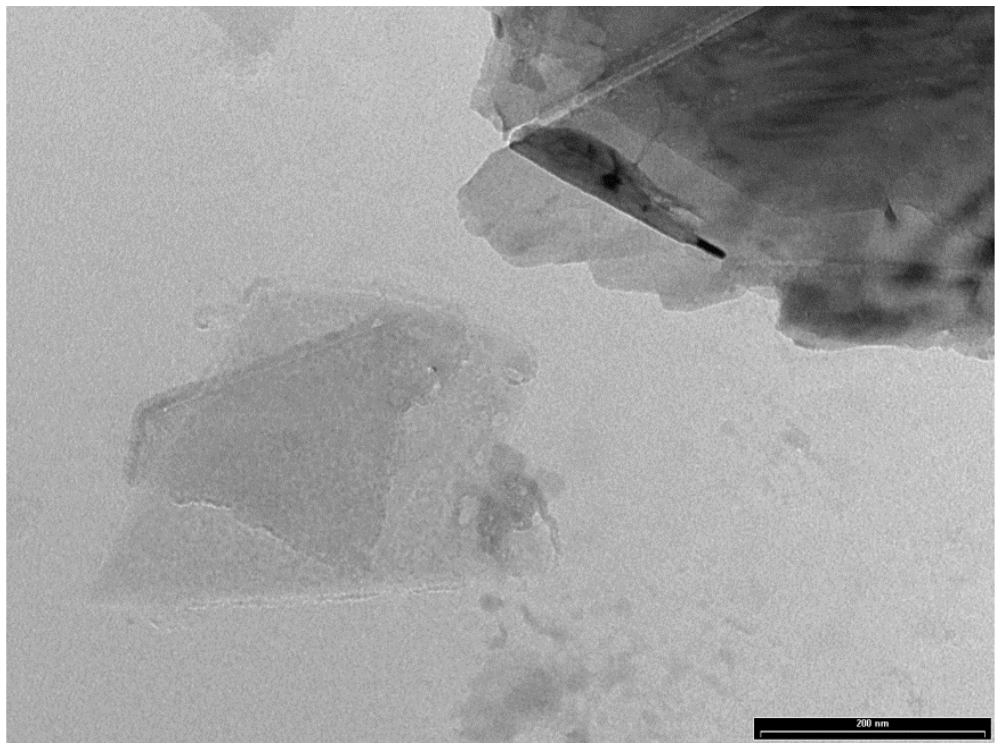

Figure 4. Left: low-resolution bright field TEM image of DMSO-exfoliated BP nanosheets using P/H$_2$O molar ratio equal to 2. Scale bar: 200 nm. Right: EDX carried out on the same sample.

The chemical quality of the BP flakes was assessed by carrying out an EDX (energy-dispersive X-ray spectroscopy) analysis (Figure S9). We observed only the peak corresponding to P element, and to copper, due to the support.
AFM analysis on the same sample confirmed the nanosheets are homogeneous on a large scale with heights ranging from few nanometers to ten nanometers. Figure 5 shows the height map and the relative height profile. Considering a layer-to-layer spacing of 0.53 nm, the thickness of ~2 nm shown in Figure 5 corresponds to four atomic layers of BP. AFM and Raman measurements were carried out under air, indicating the sample does not degrade during the time of measurements.

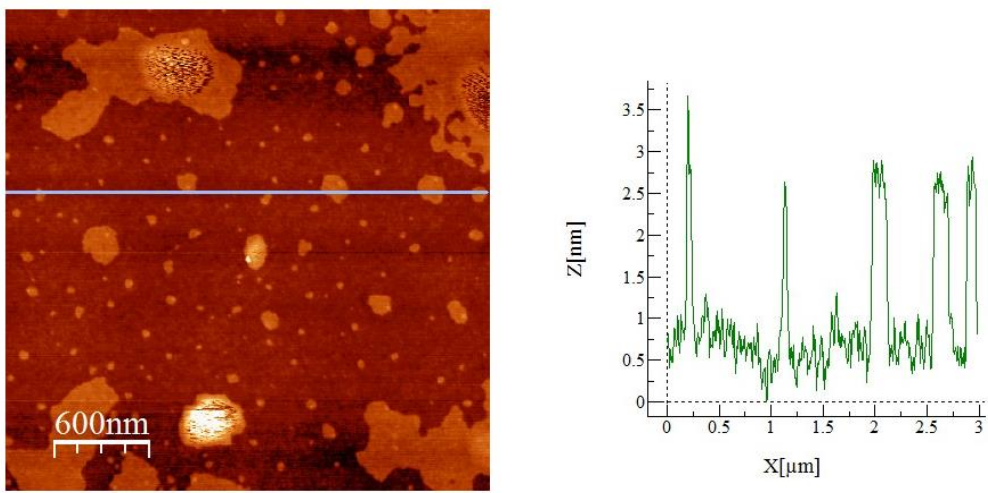

Figure 5. Left: AFM image obtained from drop casted BP flakes of a sample having P/H$_2$O molar ratio equal to 2, scale bar = 600 nm. Right: height profile map.

$^{31}$P and $^{31}$P{$^{1}$H} NMR studies were carried out at regular intervals of time during the sonication and showed the presence in solution of only H$_3$PO$_3$ and H$_3$PO$_4$ as side-products (see Figure S8). The degradation of phosphorene to P-oxyacids in this range of P/H$_2$O molar ratio was a minor process and it was not accompanied by formation of polyphosphates, as occurred in Test A. Additionally, we verified that the dispersion of phosphorene in DMSO was stable at room temperature under nitrogen and in the dark for at least one week, while in air the stability was reduced to few hours.

Powder X-ray diffraction (PXRD) of the dark grey solid which settled down after centrifugation was also carried out to confirm the nature of the sample and its purity. As shown in Figure 6, only few peaks were observed, which correspond to the 020, 040 and 060 reflections of BP, while any other peaks were absent or extremely weak. This again confirms the purity of the sample because no other crystalline phases were detected, which is a consequence of the preferred orientation of black phosphorus flakes on the flat support showing that phosphorene layers form a film with the 2D (010)-planes parallel to the support surface.

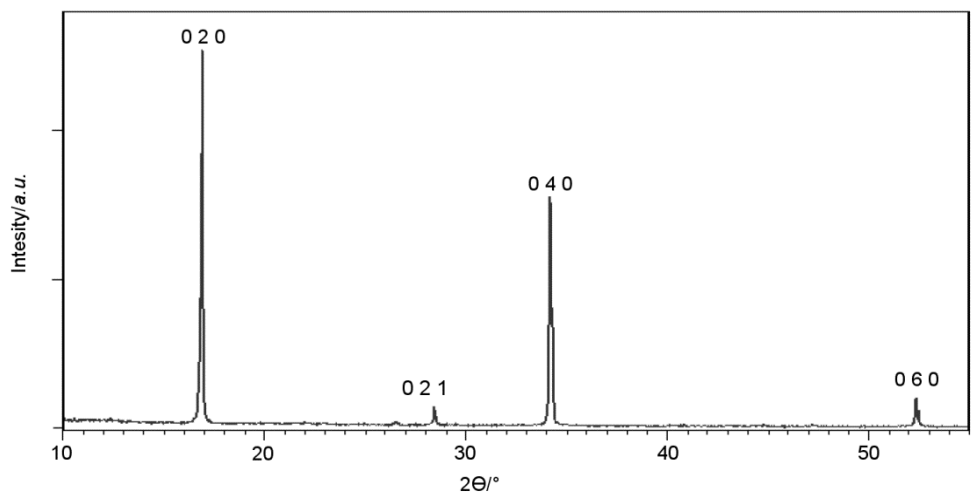

Figure 6. X-ray powder pattern of exfoliated black phosphorus.

The Raman spectrum is shown in Figure 7. It is averaged over the area shown in the inset, where the area of the Raman peak at ~463 cm$^{-1}$ is shown in arbitrary units (scale bar: 4 μm). Raman mapping revealed the three typical bands of phosphorene, centered at 363 cm$^{-1}$, 440 cm$^{-1}$ and 468 cm$^{-1}$ attributed to $A^1_g$, $B_{2g}$, and $A^2_g$ vibrational modes, respectively. In keeping with the crystalline nature of the sample, the average ratio between the area of the $A^2_g$ and $A^1_g$ peaks was ~2.4, confirming the multilayer nature of our samples.[17]

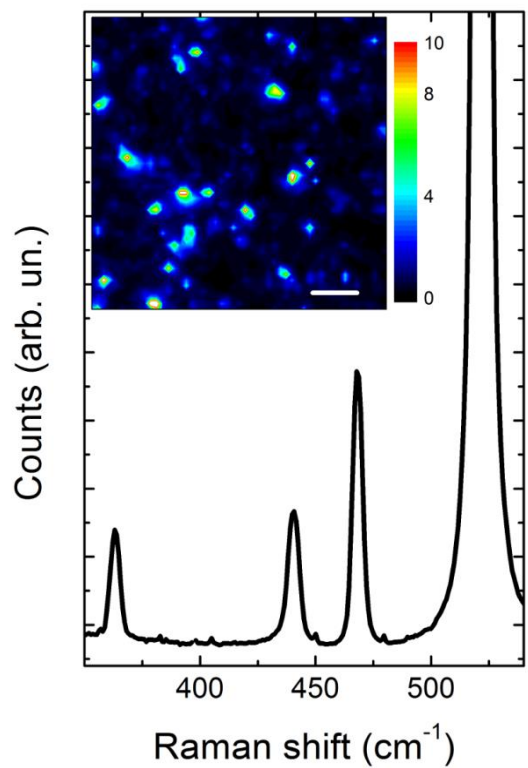

Figure 7. Raman spectrum of solvent-exfoliated BP nanosheets. A large peak from the Si substrate is found at 521 cm$^{-1}$.

Test C) 0.3 < P/H$_2$O < 1.4

Running the same sonication experiment in DMSO with a larger amount of water (P/H$_2$O molar ratio equal to 0.7) in the same way did not afford any homogeneous sample in the optical microscopy, while AFM and Raman showed piles of black phosphorus flakes together with large empty areas, see Figure 8. The thickness of the flakes was higher than in B) and ranged from 30 nm to about 200 nm. We deduce that under these experimental conditions the liquid exfoliation was least effective.

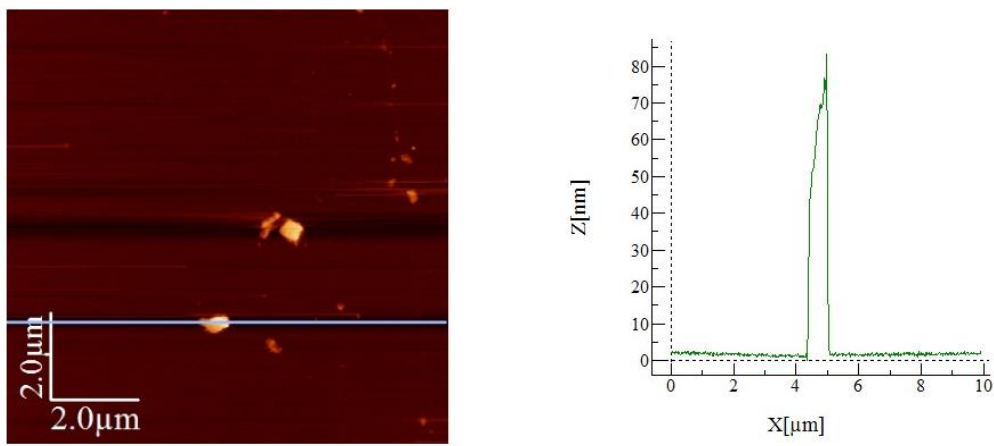

Figure 8. Left: AFM image from drop casted BP flakes of a sample with P/H$_2$O molar ratio equal to 0.7, scale bar = 2.0 μm. Right: height profile map.

SEM and TEM measurements showed BP flakes with jagged borders and irregular shape, see Figure 9 and S10, respectively. Moreover, the lateral dimension of the layers was far from being homogeneous, ranging from small size, *i.e.* 120 nm to larger size, 2300 nm.

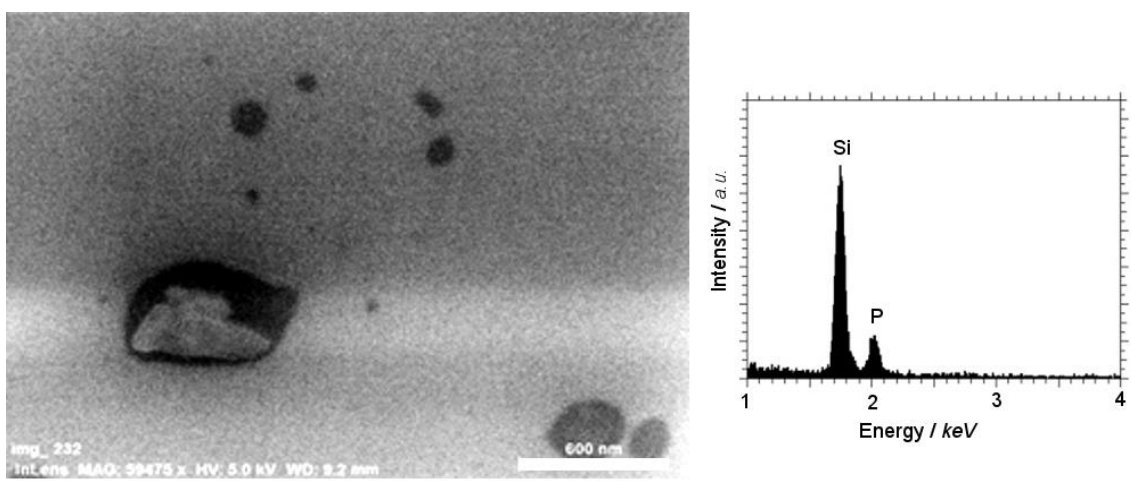

Figure 9. Left: SEM image, scale bar: 600 nm. Sample with P/H$_2$O molar ratio equal to 0.7. Right: EDX on BP flakes.

During SEM measurements, an EDX analysis performed on the BP flakes (see Figure 9, right side), detected phosphorus the silicon peak being due to the support, meaning the high purity of our sample and the absence of oxidation on BP nanosheets.

$^{31}$P and $^{31}$P{$^1$H} NMR were again carried out at regular intervals of time during the sonication and showed again the presence in solution of only H$_3$PO$_3$ and H$_3$PO$_4$ as side-

products (Figure S12). As for case B, the degradation of phosphorene P-acids in this range of P/H$_2$O molar ratio, was a secondary process not affording any phosphates.

In summary, for the first time the role of water content in the process of liquid exfoliation of BP carried out in DMSO was experimentally studied. Combining electron microscopy techniques, such as TEM and AFM, with NMR spectroscopy and ESI MS analysis, the dramatic influence of water emerged, which can stir the process towards the formation of high quality BP nanosheets or to undesired polyphosphorus molecular species. If the molar ratio P/H$_2$O is greater than 15, degradation products are obtained particularly trimetaphosphate which are likely produced from oxidation by DMSO. In contrast, if the molar ratio P/H$_2$O is in the range between 1.5 and 14, high quality flakes having lateral dimension ranging from 400 nm to 600 nm and thickness between 2 nm and 10 nm are obtained. Finally, when the molar ratio P/H$_2$O is in the range between 1.4 and 0.3, the sample looks inhomogeneous, and larger flakes with thickness ranging from 10 nm to 200 nm and lateral dimensions up to 1600 nm are obtained.

**Supporting Information**

Supporting Information is available from the Wiley Online Library or from the authors.

**Acknowledgments**

The authors thank the European Research Council (ERC) under the European Union's Horizon 2020 research and innovation programme (Grant Agreement Nº 670173) for funding the project **PHOSFUN** "*Phosphorene functionalization: a new platform for advanced multifunctional materials*" through an ERC Advanced Grant. Thanks are due also to ECRF (FIRENZE HYDROLAB2) and CNR (National Research Council, Rome, Italy, through the projects PREMIALE 2011 and 2012) for financial support and for a research grant to MSR. Dr. Daniele Ercolani (NEST, Istituto Nanoscienze-CNR and Scuola Normale Superiore, Pisa, Italy) is thanked for running SEM and EDX measurements.

**Experimental Section**

All chemicals were reagent grade and, unless otherwise stated, used as received by commercial suppliers. All reactions were carried out under a pure nitrogen atmosphere by using standard Schlenk or glove-box techniques with freshly distilled and oxygen-free solvents. Red phosphorus 99.99%, $SnI_4$, Au foil, Tin wire and anhydrous DMSO were supplied from Aldrich and used without further purification. DMSO-$d_6$ (Aldrich) was pre-treated with three freeze-thaw pump cycles before use. Black phosphorus was prepared as described in the literature[10] and stored in a vial under nitrogen and in the dark. The quality of the obtained BP crystals was checked by optical microscopy and by X-ray powder diffraction.

The sonication bath was an Elmasonic P70H, operating at 37 kHz frequency.

Samples were first imaged by an optical microscope. Immediately after, they were transferred to an AFM for high-resolution imaging. AFM scans were performed by a Dimension Icon (Bruker) instrument configured in "Scan-Asyst" mode, where topography and in-phase signal images were simultaneously acquired. Between measurements the samples were kept under a nitrogen atmosphere.

$^1$H, $^{31}$P and $^{31}$P{$^1$H} NMR were routinely recorded using a Bruker Avance II 300 spectrometer. Peak positions are relative to tetramethylsilane and were calibrated against the residual solvent resonance ($^1$H). Chemical shifts for $^{31}$P NMR spectra were measured relative to external 85 % $H_3PO_4$ with downfield values reported as positive.

IR spectra were recorded with a Spectrum BX II Perkin-Elmer spectrometer. Samples were dried under vacuum warming up at *ca.* 50 °C to eliminate DMSO and the resulting viscous oil was deposited on KBr platelets.

Raman spectra were collected with a custom-built micro-spectrometer. The 488-nm line from an Ar laser was focused on the sample with a 63x objective with a numerical aperture of 0.9. The power on the sample was 0.3 mW. Light emitted by the sample was dispersed by a HR550 Jobin-Yvon monochromator and the spectrum collected by a Synapse CCD.

TEM: studies were carried out using a Philips instrument operating at an accelerating voltage of 100 kV.

SEM: Zeiss Ultraplus field emission SEM operated at an accelerating voltage of 5 keV.

EDX: Bruker Quantax 800 EDX having an energy resolution of ~130 eV, and operated with a electron beam energy of 5kV.

ESI-MS spectra were recorded by direct introduction of the samples at 7 μl/min flow rate in an LTQ-Orbitrap high-resolution mass spectrometer (Thermo, San Jose, CA, USA), equipped with a conventional ESI source. The working conditions comprised the following: spray voltage 5 kV, capillary voltage 35 V, capillary temperature 275 °C, tube lens 50 V. The sheath, auxiliary and the sweep gases were set, respectively, at 8 (arbitrary units), 2 (arbitrary units) and 2 (arbitrary units). For acquisition, Xcalibur 2.0. software (Thermo) and a nominal resolution (at m/z 400) of 100,000 were used.

The samples were prepared by dilution with acetonitrile of the reaction mixture in DMSO and the analysis were carried out in negative ion mode.

X-ray powder diffraction (XRD) data were collected with an X'Pert PRO diffractometer with Cu-K$\alpha$ radiation ($\lambda$ = 1.5418 Å). XRD spectra were acquired at room temperature with a PANalytical X'PERT PRO diffractometer, employing Cu $K_\alpha$ radiation ($\lambda$ = 1.54187 Å) and a parabolic MPD-mirror for Cu radiation. The diagrams were acquired in a 2$\theta$ range between 10.0° and 5.0°, using a continuous scan mode with an acquisition step size of 0.0263° and a counting time of 54.6 s.

Procedure for black phosphorus exfoliation

**Test A**: molar ratio (P/H$_2$O) = 15.

A 5 mm NMR test tube was charged with 0.6 mL of DMSO-$d_6$ (H$_2$O ~ 75 ppm) and crystalline black phosphorus (4.0 mg, 0.13 mmol) (P/H$_2$O = 15). The mixture was freeze-pump-thaw degassed three times under nitrogen, before being sonicated (37 kHz) for 20 hours keeping the temperature fixed at 30 ºC.

**Test B**: molar ratio (P/H$_2$O) = 2.

A 5 mm NMR test tube was charged with 0.6 mL of DMSO-$d_6$ (H$_2$O = 75 ppm), BP crystals (4.0 mg, 0.13 mmol) and deoxygenated water (1.0 µL, 0.056 mmol) (P/H$_2$O = 2.0). The procedure was the same as Test A. Under these conditions, the suspension remained stable for at least one week.

**Test C**: molar ratio (P/H$_2$O) = 0.7.

A 5 mm NMR test tube was charged with 0.6 mL of DMSO-$d_6$ (H$_2$O = 75 ppm), BP crystals (4.0 mg, 0.13 mmol) and deoxygenated water (3.0 µL, 0.167 mmol) (P/H$_2$O = 0.7). The mixture was treated as in Test A. Under these conditions, the suspension remained stable at least four weeks.

# Supporting Information

# The role of water in the preparation and stabilization of high quality phosphorene flakes

*Manuel Serrano-Ruiz,[a] Maria Caporali,[a,]* Andrea Ienco,[a] Vincenzo Piazza,[b] Stefan Heun,[c] Maurizio Peruzzini[a,]**

Test A) P/$H_2O$ molar ratio equal to 15.0.

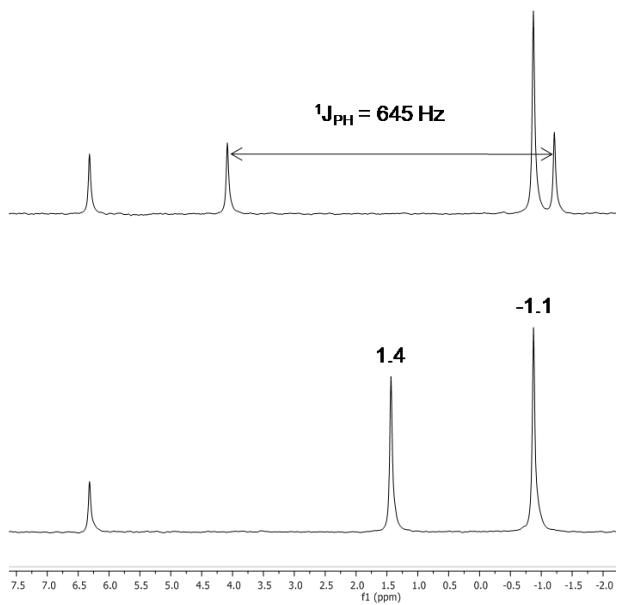

Figure S1. Up: $^{31}$P NMR and down: $^{31}$P {$^1$H} NMR (DMSO-$d_6$, T = 25°C, 121.49 MHz) spectrum of a sample of Test A after 4 hours of sonication.

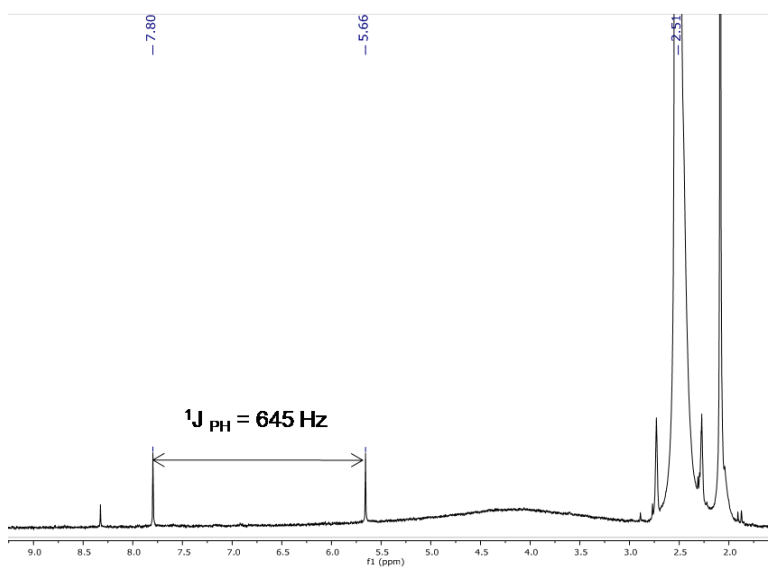

Figure S2. $^1$H NMR spectrum (DMSO-$d_6$, T = 25°C, 300.13 MHz) after 4 hours of sonication.

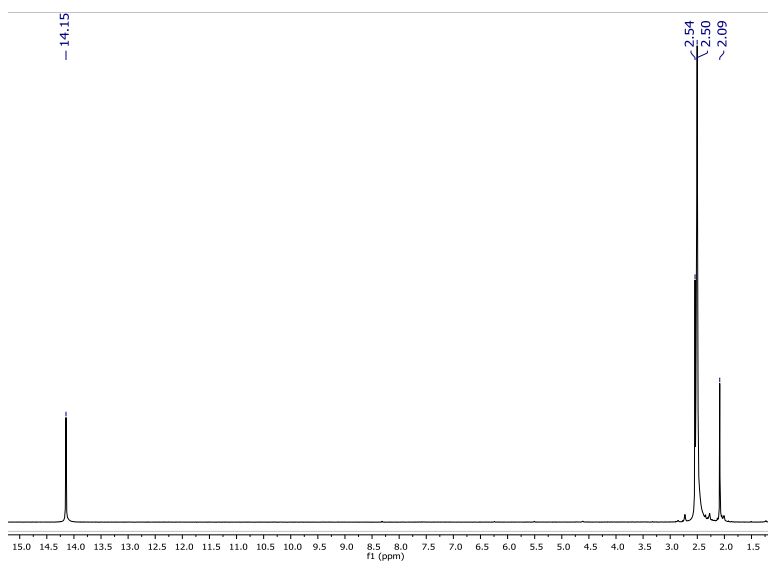

Figure S3. $^1$H NMR spectrum (DMSO-d$_6$, T = 25°C, 300.13 MHz) after 20 hours of sonication.

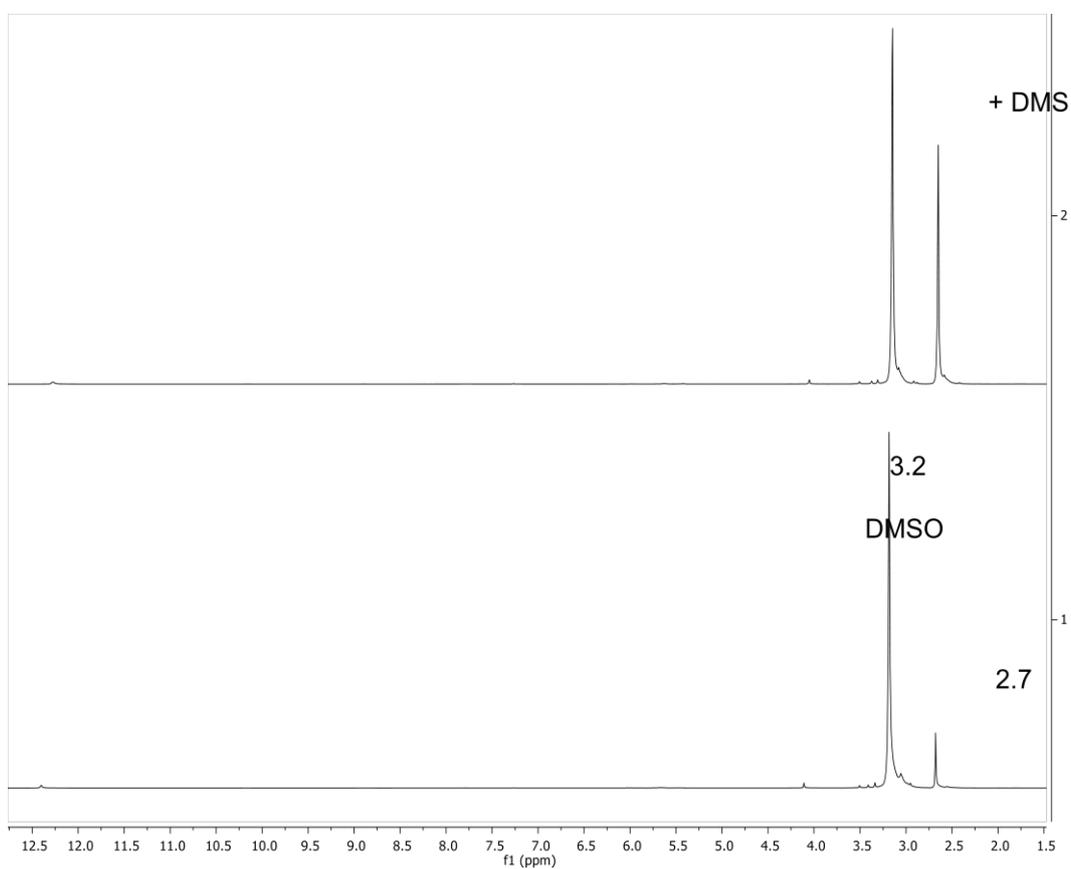

Figure S4. $^1$H NMR (DMSO, T = 25°C, 300.13 MHz, capillary C$_6$D$_6$). Down: reaction mixture before adding [S(CH$_3$)$_2$]; top: after adding [S(CH$_3$)$_2$].

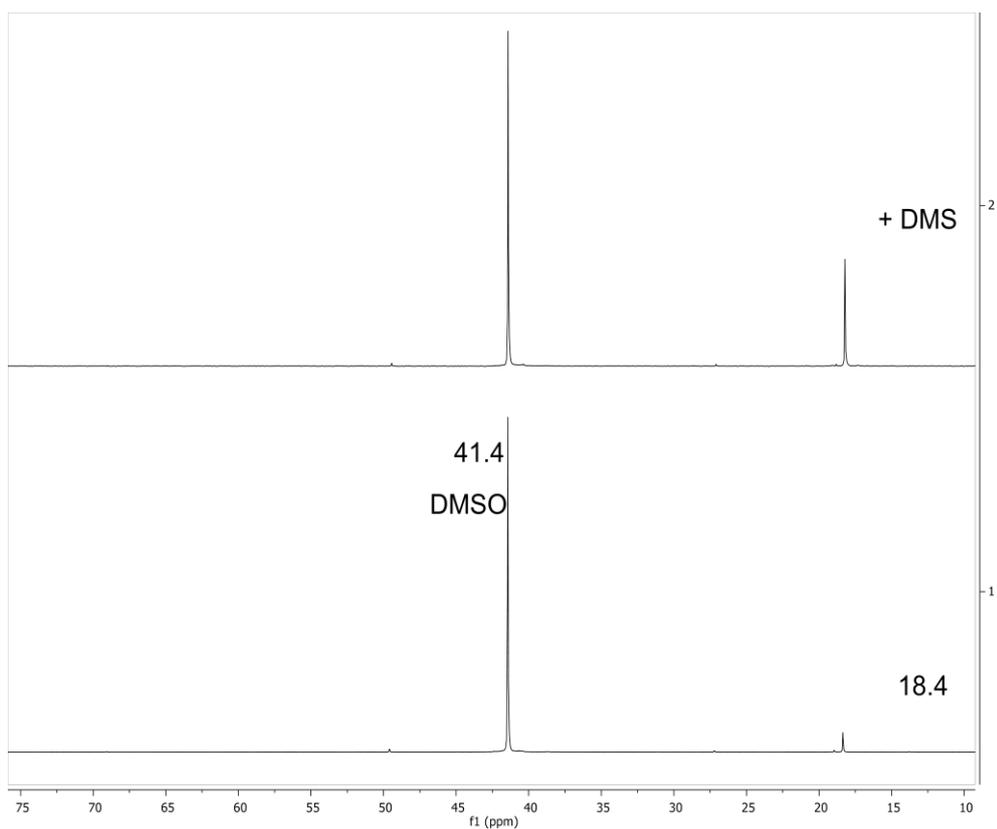

Figure S5. $^{13}$C NMR (DMSO, T = 25°C, 75.5 MHz, capillary $C_6D_6$). Down: reaction mixture before adding [S(CH$_3$)$_2$]; top: after adding [S(CH$_3$)$_2$].

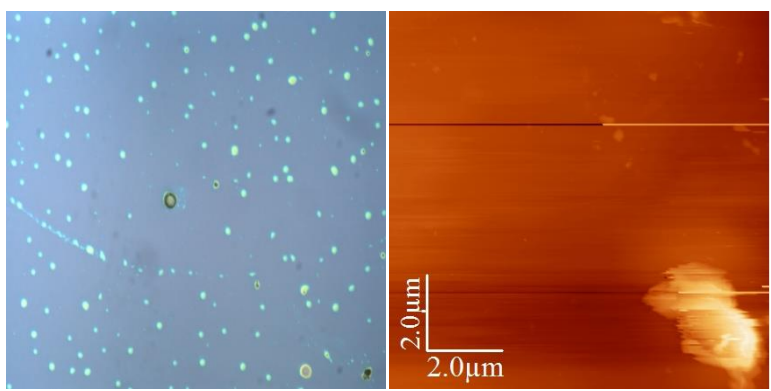

Figure S6. Left: optical microscopy image; Right: AFM of a sample of Test A after 20 hours of sonication, scale bar 2.0 μm.

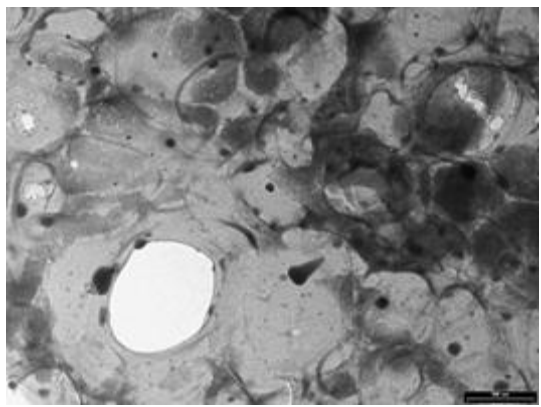

Figure S7. Bright field TEM image on holey carbon Cu grid of the sample in DMSO where P/H$_2$O molar ratio equal to 15. Scale bar: 500 nm.

Test B) P/H$_2$O molar ratio equal to 2.0.

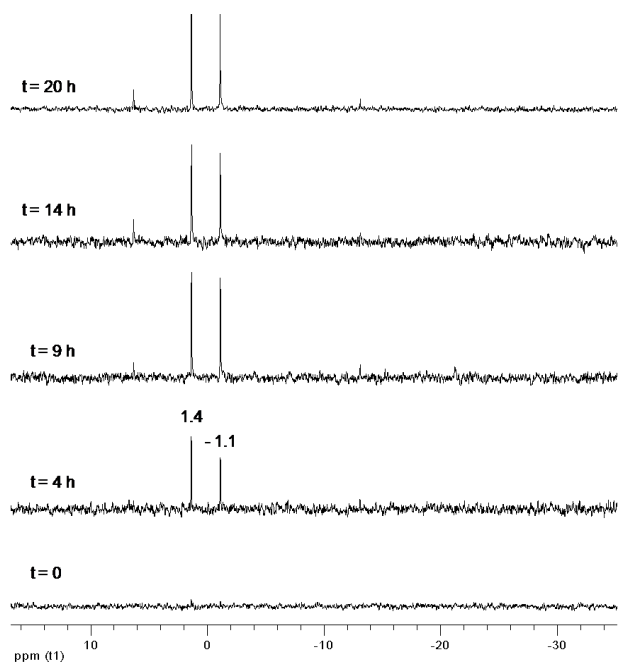

Figure S8. $^{31}$P{$^1$H} NMR spectra (DMSO-d$_6$, T = 25°C, 121.49 MHz) of a sample of Test B, P/H2O molar ratio equal to 2, during sonication.

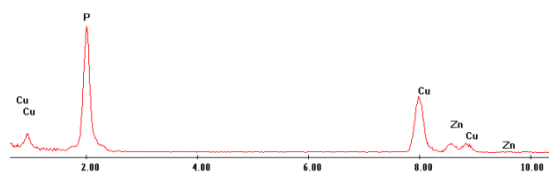

Figure S9. EDX carried out on the sample of Test B, P/H2O molar ratio equal to 2, after 20 h of sonication.

Test C) P/H$_2$O molar ratio equal to 0.7.

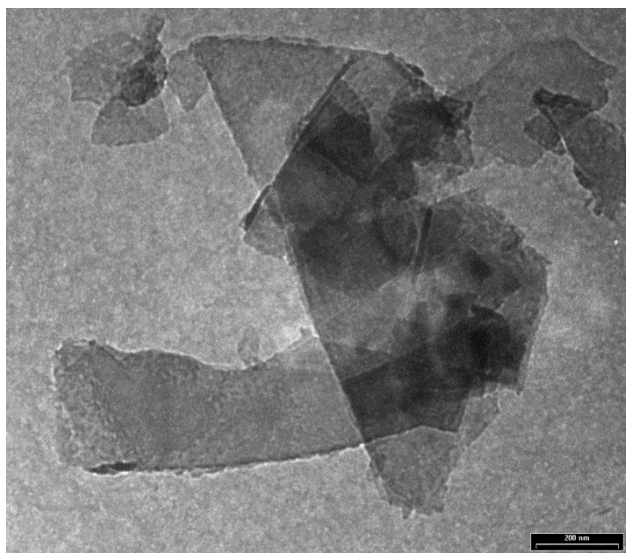

Figure S10. Bright field TEM image of DMSO-exfoliated BP nanosheets. Scale bar: 200 nm.

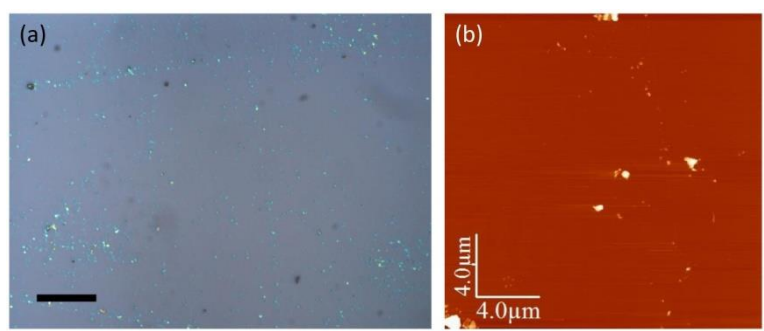

Figure S11. Left: Optical microscopy image, 144 μm x 108 μm, scale bar 30 μm; right: AFM image of a sample with P/H$_2$O molar ratio equal to 0.7.

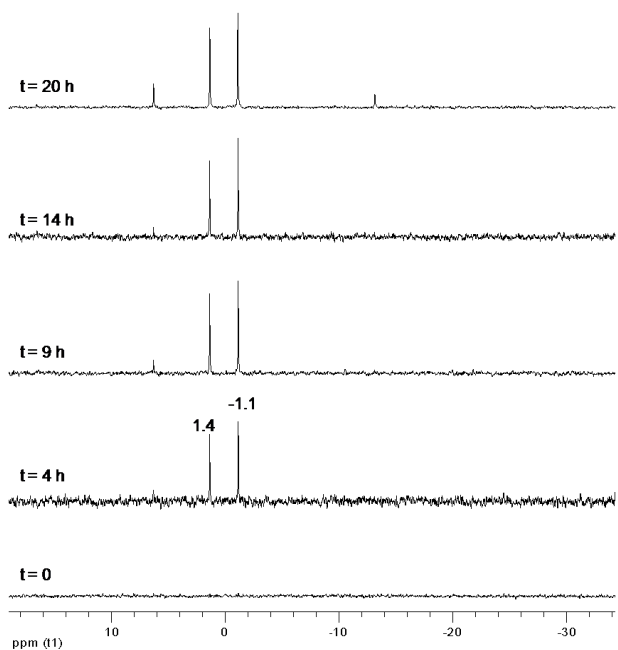

Figure S12. $^{31}P\{^1H\}$ NMR spectra (DMSO-$d_6$, T = 25°C, 121.49 MHz) measured during the sonication.

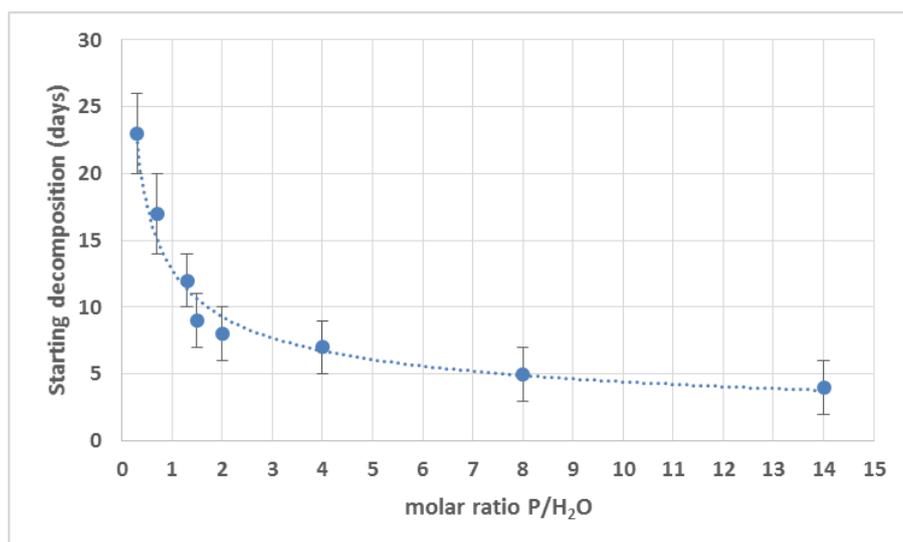

Figure S13. Starting of the exfoliated BP decomposition plotted against the P/H$_2$O molar ratio (as shown by the appearance of the -24.9 ppm signal in the $^{31}P\{^1H\}$ NMR)

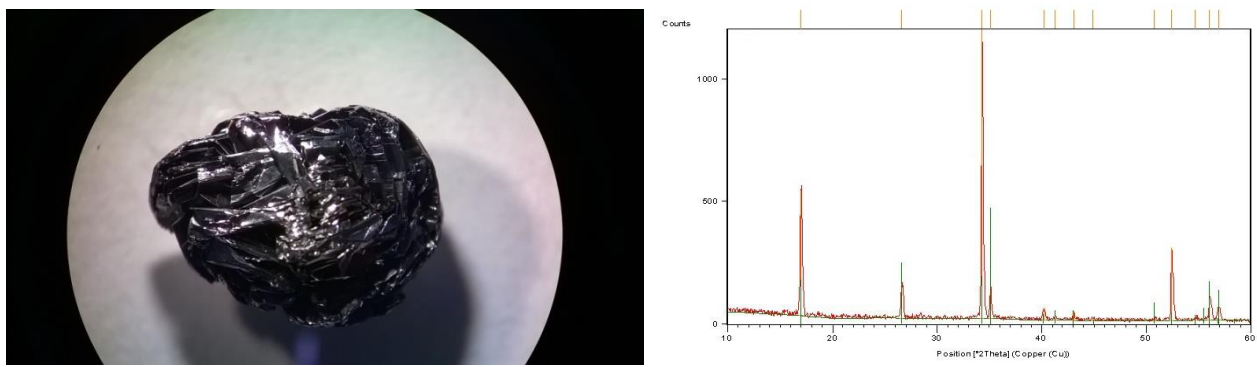

Figure S14. Picture of a BP crystal obtained after the synthesis (left); X-Ray powder diffraction of BP crystals (right).